\magnification=1200
\def\ni{\noindent}

\def\npb{{\it Nucl.~Phys. B}}

\def\plb{{\it Phys.~Lett. B}}

\def\prs{{\it Proc. Roy. Soc. }}

$\,$\hfill SWAT00/278
\medskip
\ni{\bf The Quantisation of Charges}

\bigskip
\ni {\sl David I Olive, University of Wales Swansea}
\bigskip
\ni{\it Invited lecture at the Symposium
 \lq\lq One hundred years of h", Pavia, September 2000}
\bigskip

{\it The question as to whether the integrality
 of the spectrum of observed electric charges is due to a quantum effect
has fascinated theoretical physicists throughout
the last century. It leads to unanswered questions
 at the heart of quantum field
and its role as a framework for particle physics.}
\bigskip
As you have heard, it is one hundred years ago 
since Planck introduced his famous constant
when he proposed that the energy of components of black
body radiation of angular frequency $\omega$ occurred in
discrete quanta:
$$E=\hbar\omega,\eqno(1)$$
in modern notation. This revolutionary proposal
 resolved problems with the intensity spectrum of the radiation,
and led eventually to the development of quantum mechanics,
a theory in which the constant $\hbar$ plays a pivotal role.
According to this new theory, because the electromagnetic field responsible
for the black body radiation is dynamical, it
 ought to be subjected to the principles of quantisation.
These provide a well-defined technical procedure and yield the result
that this field had quantum excitations that
should be identified as particles, in this case the photon [1].
Then it follows that the energy of a photon associated with an electromagnetic
wave of frequency $\omega$ is indeed given by the above expression (1).

This correspondence between particles and fields has been very
much extended, for example, to the electron and its wave function, 
regarded as a field.
Indeed similar results apply to all the known particles discovered
in high energy particle accelerators and in cosmic rays. Basic
theory in modern day particle physics theory exploits this formalism.

There is another more obvious pattern observed first in
atomic physics and applicable to all particle physics. 
Atoms (unless they are ionised)
are electrically neutral despite the fact that a number of electrons
orbit around their nuclear core. It follows that the atomic
nucleus must carry an electric charge that exactly cancels that
of the electrons and hence equals a negative integer times
 the electron charge. In fact the nucleus is composed
of neutrons which are chargeless and protons carrying an electric
charge equal and opposite to that of
the electron. Indeed all observed particles, such as the muon, the pions,
and so on,
 carry an electric
charge that is an integer multiple of the electron charge
$$q=nq_0,\qquad n=0,\pm1,\pm2,\dots \eqno(2)$$

This integrality property of electric charges in appropriate units
 is one of the most striking
features of particle physics. Planck's constant is missing
from the relation (2) yet it looks so much like a quantum
phenomenon that it is natural to wonder whether there is
some way to deduce it from the principles of quantum mechanics, just as
(1) has been.
This question leads to an exciting story that I want to tell you about. 
It is a
 story still unfinished, one that probes to the very heart of quantum field theory 
and its concepts,  leading to the conclusion that
there is still much to learn about the quantum theory,
particularly when it is applied to the most fundamental
questions, namely of finding a relativistic description
of unified particle interactions.

With the electron charge $q_0$ it is useful
to construct a dimensionless number that is important in
quantum electrodynamics. It is called the \lq\lq fine structure
constant":
$$\alpha_{{FINE\atop STRUCTURE}}={q_0^2\over2\pi\hbar c}\sim{1\over137}.\eqno(3)$$
Its experimental value is small as indicated. 
In the early days of quantum theory it was hoped that this number
could be determined. Many crazy but unsuccessful ideas were advanced to this
end, for example by Eddington, and a more recent group theoretic
idea was described at this meeting. But one of these
seemingly crazy attempts has survived, remaining tantalising, influential
and pervasive ever since. This is the proposal that Dirac made in 1931 [2].

He considered the possibility that  somewhere in nature there existed a magnetic
monopole, a new sort of particle carrying a magnetic charge, $g$,
and not yet seen. It was known previously
that there was no problem in modifying Maxwell's equations to accommodate
this new sort of charge. However there could be a problem with the
newly discovered quantum theory and indeed Dirac found that, 
in general, there was. 
This difficulty could be circumvented
only if the following relation held for any pair of particles
with electric charge $q$ and magnetic charge $g$, respectively:
$$qg=2\pi\hbar c n,\qquad n=0,\pm1,\pm2,\dots\eqno(4)$$
The argument leading to this \lq\lq Dirac quantisation condition"
has been steadily refined and it is clear that
it depends upon very little, just the idea that the electrically
charged particles possess quantum wave functions that can be pieced together
in a consistent way, making careful use of the gauge principle. No particular equations
of motion are needed nor other details.

Let $q_0$ be the observed electric charge of smallest magnitude (and hence
the electron charge) and $g_0$ the possible magnetic charge of smallest
non-vanishing magnitude.
Then it is reasonable that these satisfy (4) with $n=1$ (though $n$ could
have a higher value):
$$q_0g_0=2\pi\hbar c \eqno(5)$$
and that
$$q=nq_0,\qquad n=0,\pm1,\pm2\dots\eqno(6a)$$
$$g=mg_0, \qquad m=0,\pm1,\pm2\dots\eqno(6b)$$
Thus the integrality pattern  of electric charge does follow
from a quantum principle
so it is now seen to be perfectly legitimate to talk of charge quantisation.
 But this is at a price, namely
 that there has to be a   magnetic charge somewhere
in the universe, not yet observed.
Nevertheless Dirac found this result gratifying and it is still true today, after
seventy years development in quantum field theory,
 that this simple argument
remains the best explanation of the integrality relations $(2)$ for electric charge.

Dirac's original intention was to determine $\alpha$, (3), numerically.
 Instead he found the relation following from (5):
$$\left({q_0^2\over2\pi\hbar c}\right)\left({g_0^2\over2\pi\hbar c}\right)
=\left({q_0g_0\over2\pi\hbar c}\right)^2=1.\eqno(7)$$
So, according to (3), 
 the magnetic \lq\lq fine structure constant", $g_0^2/2\pi\hbar c$,
is approximately $137$, significantly larger than unity and
hence what is called \lq\lq strong", in contrast with (3), the electric
fine structure constant, which, 
being much less than unity, is said to be weak.

Until recently the main computational tool in quantum field theory
was what is known as perturbation theory and it relied on the dimensionless
coupling, such as the electric fine structure constant,  (3), being smaller than unity.
 The situation just described,
with electric and magnetic charges, is remarkably symmetrical 
between the two, despite the disparity in their dimensionless magnitudes.
Maybe a quantum field theory could be constructed in which this symmetry 
between the two roles is maintained. If so, perturbative calculations
could be performed exploiting the smallness of the electric coupling, (3)
and might yield information about the magnetic features despite
their strong coupling. This is a very tantalising prospect
and, surprisingly, it seems to be true, at least in suitable cases.

The onus is of course to find such a quantum field theory and it must
incorporate much additional structural information about the charged particles
such as the ones we have mentioned. In particular their masses
 and internal structure will have to come into play.

In the  last forty years or so the subject of quantum field theory has 
burgeoned under the stimulation of new ideas (confirmed in part by
experimental data) for describing all the elementary particle interactions in a
unified way. The simplest concept of unification is to extend Maxwell's equations
to a more complicated, non-linear system associated with more 
charge operators
as well as  the electromagnetic one, denoted $Q$. These charges
form a closed algebra under commutation, called a Lie algebra and can be
exponentiated to yield a Lie group,
 called the gauge or Yang-Mills group [3].
The nonlinearity of the modified Maxwell equations is reflected
in the fact that this group is non-abelian,
 namely that the order of multiplication within the group matters.

The simplest new possibility is to have just three charges in all, $T_1$, $T_2$
and $T_3$ say. Then the  only possible non-abelian group consists
of rotations in a three-dimensional space. This has to be an internal
space rather than the real space in which we live. But, as in the
latter case, the three charges must satisfy the angular
 momentum commutation relations
$$[T_1,T_2]=iT_3\qquad\hbox{etc.}\eqno(8)$$
The charge operator $Q$ must be one of these, or, more precisely,
must be a linear combination of them. Thus it has to be
 an angular momentum operator
about some direction in the internal three-space. It could, for example
be proportional to $T_3$, except that it would not be satisfactory
to select a particular direction by such a fiat as 
the choice would not be symmetrical in the sense of the gauge symmetry
which is the new guiding principle.

To accommodate this principle it is necessary to incorporate three
new scalar fields $(\phi_1,\phi_2,\phi_3)$ that form a vector in the
three dimensional space thereby creating a signpost that designates 
the electromagnetic  direction (at each of the 
points of space-time for which they are defined).

Then the candidate electric charge operator is proportional to the charge
selected thus:
$$Q\sim \underline{\phi}.\underline{T}.\eqno(9)$$
But since the charges are to be non-zero,
 this is not allowed to vanish, even in the vacuum. This
means that the fields $\phi$ cannot vanish there, unlike normal fields.
Consequently they are what is known as Higgs fields [4], more realistic examples
of which
 are being sought with such urgency at CERN.

So the quantum field theory under consideration has now become
  what is called a spontaneously
broken gauge theory comprising both gauge and Higgs fields. The Higgs
fields will perform their original dedicated role and contribute
 mass for the gauge particles.
The photon will remain massless, as is desirable, while
 the other two, called $W^{\pm}$,
antiparticles of each other, will have a specific mass 
(as is also desirable)
 and function
as intermediate vector bosons for an (oversimplified) toy model of weak
or radioactive interactions. This model is known as the
 Georgi-Glashow model [5].

The three charge operators $T$ can be thought of as matrices
and this therefore gives a matrix for $Q$, by (9). Its eigenvalues are 
proportional to the electric charges carried by the various
corresponding particles of the theory and are automatically quantised, by 
the quantum theory of angular momentum applied to the 
commutation relations  (8). Thus charge quantisation
is apparently achieved painlessly without a magnetic monopole in sight.

Now comes a surprise due to two features of the
 theory that have already been mentioned, first,  that the equations
of motion are highly non-linear and, secondly,
 that the Higgs field, $\phi$,
is able to \lq\lq swivel around" from point to point of space-time
whilst remaining in its ground state or vacuum, and maintaining eigenvalues
or physical electric charges that do not vary from point to point.
 The result is that the equations possess classical solutions
 that can be regarded
as \lq\lq solitons", so that the energy density remains
 localised in regions of space. Indeed these are analogues of the
solitons of sine-Gordon theory familiar in space-times of two
dimensions. Those simple solitons can be demonstrated with a toy made of pins and rubber
as one of the speakers showed, whereas the present ones can only be demonstrated
by explicit analytic solutions that make clear
 that there are no sorts of singularity. What happens is that
 in order for a classical
solution to have finite energy, the scalar Higgs 
field evaluated at large distances in space, 
that is on the two-sphere at infinity, must  take values
in its ground state or vacuum which itself forms a two-sphere
in the three dimensional internal space. Thus the asymptotic
 Higgs field
provides a map between a pair of two-spheres and these
maps are characterised by an integer, called a degree, that
furnishes a higher dimensional generalisation of
 the winding number that classifies maps between two
one-spheres, or circles. This degree cannot be changed 
by the supply of any finite amount of energy and so provides a sort of topological 
quantum number for the soliton configurations. Something
 very similar happens for the sine-Gordon solitons.
The single soliton (or anti-soliton) corresponds to the smallest value
of this degree, $\pm1$, when one sphere is transported bodily onto the other
with or without reflection, with the asymptotic 
Higgs' field swivelling appropriately.
 The soliton stability is guaranteed by the the conservation
of the topological quantum number.

This topological number ought to
have a more direct physical interpretation and indeed it does,
for it is proportional to the magnetic charge defined as a flux out of the
asymptotic sphere  of the magnetic components of the Maxwell
field in the unified theory. These are the famous
 results that 't Hooft and Polyakov found independently in 1974 [6].

This magnetic charge automatically satisfies the Dirac quantisation
condition (4), even though quantum theory has not been explicitly invoked.
Furthermore the associated monopole particle, and its antiparticle,
 $M^{\pm}$, are endowed with a mass
whose expression tantalisingly resembles that possessed by the $W^{\pm}$ gauge particles,
 by virtue of the Higgs mechanism mentioned earlier.

The first conclusion is that Dirac's explanation of charge quantisation is triumphantly
vindicated. At first sight it seemed as if the idea
of unification provided an alternative explanation, avoiding monopoles,
but this was illusory as 
 magnetic monopoles were
indeed lurking hidden in the theory, disguised as solitons.

This raises an important conceptual point. 
The magnetic monopole here has been treated
as bona fide particle even though it arose as a soliton,
namely as a solution to the classical equations of motion.
It therefore appears to have a different status from the
\lq\lq Planckian particles" considered hitherto and discussed at the beginning
of the lecture.  These arose as quantum excitations
 of the original fields of the initial formulation of the theory,
products of the quantisation procedures applied to these dynamical variables
(fields).

The point is that once quantum theory is fully applied, as it must be,
 the solitons too will appear as particles in the
 only real sense that concept has, namely that
 enunciated by Wigner in 1939, 
technically as irreducible representations of the Poincar\'e group of
space-time transformations [7]. The particles appearing by virtue of 
the two mechanisms,
quantum excitation and soliton,
must therefore have an entirely equivalent status. The apparent difference
is just an artefact of the way the quantum field theory has been formulated.
This point was first made clear by Skyrme
 in his discussion forty years ago of the solitons of the sine-Gordon model in space-times 
of two dimensions [8]. He found that there do exist
 field operators of which the solitons
are quantum excitations. They are related to the original sine-Gordon field
by what is now recognised as a vertex operator construction.
 Furthermore, as Mandelstam showed [9],
 they satisfy
the equations of motion of what is called the massive Thirring theory. 
As a result this theory is quantum equivalent to the sine-Gordon theory.

Two questions now arise naturally,
 (1) the construction of the field operators
\lq\lq creating" the magnetic monopole solitons $M^{\pm}$, and
 (2) the identification of the equations of motion
that these fields satisfy. In space-times of the interesting dimension,
$4$, that is under consideration, unlike the easier dimension, $2$, 
the first question is still far too difficult to answer. But it is possible
to hazard a guess for the answer to the second question
 and then check some of the predicted consequences, thereby accumulating
circumstantial evidence in support.

The toy unified quantum field theory model so far developed 
displays a surprising degree of symmetry between the electric
 and magnetic fields $\underline E$ and $\underline B$,
 the electric and magnetic charges $q_0$ and $g_0$ and the corresponding
charged particles, $W^{\pm}$, the heavy gauge particles, and
$M^{\pm}$, the heavy monopoles. The way the masses of these particles
depend on their charges is surprisingly similar as are 
other common properties not described here.

This suggests that $M^{\pm}$ are very similar to $W^{\pm}$ and hence
likewise a pair of heavy gauge particles in a spontaneously broken
$SO(3)$ gauge theory, exactly like the original one but with
a strong magnetic coupling replacing the weak electric coupling and
related by (7). In this alternative description $W^{\pm}$
would now occur as soliton solutions. This was the conjecture
made by Claus Montonen and myself in 1977 [10] and subsequent developments
seem to have vindicated the idea. All the evidence is confirmatory,
at least once a suitable degree of supersymmetry is included [11]. 
Supersymmetry,  a symmetry mixing bosons and fermions, 
is easily achieved by adding a few extra fields 
and particles without changing anything so far described.
The benefit is that there are minimal quantum corrections to the
formulae mentioned and that there is a natural mechanism for the monopole
solitons to acquire the spin quanta carried by gauge particles.

Actually the evidence points to an even richer structure than this.
There are extra soliton states called dyons, with the same magnetic charges
as $M^{\pm}$ but also carrying  electric charges. There
are even dyon bound states of higher magnetic charge arising
by a totally new mechanism identified by Ashoke Sen [12].

Before explaining the ramifications of this it is important to
mention that the toy theory being developed possesses a second
dimensionless parameter, called $\theta$ because it is 
angular in nature [13], that was overlooked in early work
 and appeared innocuous anyway. This can be combined
with the fine structure constant  (3) to form
 a natural complex, dimensionless variable :
$$\tau={\theta\over 2\pi}+{2\pi i\hbar \over q_0^2},\eqno(10)$$
so that the imaginary part is simply the inverse of
the fine structure constant (3) and, hence,
intrinsically positive.

The consequence of the extra richness of the particle spectrum
of the quantum field theory just mentioned
is that it appears, as Sen suggested, that there are
really an infinite number of quantum equivalent reformulations
of the theory, not just the two already mentioned. Each
of these formulations is distinguished by the assignment
 of field operators to one particular dyon/antidyon pair
(or $W^{\pm}$ pair). The equations of motion satisfied
are always those of the supersymmetric spontaneously
broken $SO(3)$ gauge theory but the couplings $\tau$ differ.
The relation between two choices always takes the form
$$\tau\rightarrow \tau'={A\tau+B\over C\tau+D}, \qquad AD-BC=1,\eqno(11)$$
where the coefficients $A$, $B$, $C$ and $D$ are integers.

The transformations (11) form an infinite discrete group, called by
mathematicians $PSL(2,Z\!\!\!Z)$, or simply, the modular group.
It is an easy and instructive exercise to check that these transformations
always preserve the positivity of
 the fine structure constant, as they should.

That these transformations are quantum is made plain 
by the appearance of Planck's constant in (10). A particularly
dramatic way of stating the result is to say
that the one quantum field theory has an infinite number
of classical limits that are classically inequivalent.
Although in each of these classical limits
 $\hbar$ always tends to zero, there is,
of course, no paradox since
 what is held constant during the limiting process
varies from case to case.

It is helpful to illustrate by a particular choice
of the transformation (11), called $S$, given by
$$\pmatrix{A&B\cr C&D\cr}=\pmatrix{0&-1\cr 1&0\cr}.$$
By (11), this transformation, $S$,   sends $\tau$ to $-1/\tau$. If $\theta$ vanishes this
means that we have recovered equation (7) (with $g_0$ replaced by $q_0'$).
This was the original transformation between weak and strong coupling.

An interesting historical aside is that a transformation like this
was discovered as early as 1941, by Kramers and Wannier [14], in what
appeared to be a quite different physical context. They showed
that there was a symmetry of the partition function of the Ising model
in space of two dimensions with respect to interchange
 of high and low temperatures. They argued that
 the fixed point of this transformation had to be the critical temperature
at which a phase transition occurred. This
was one of the first significant results in the theory of phase transitions.

We have come to the end of the story for now, and I want to sum up and add
some broad conclusions. The main point is that the question
of charge quantisation has proven
 unexpectedly deep and has led quantum theory
into unexplored territory. A certain class of quantum field theories,
of which one example was developed above, display a surprisingly
rich spectrum of particle states characterised by quantum numbers
that were not inserted initially. As a consequence
there are many quantum equivalent reformulations of the
one basic theory appearing quite different classically. The reformulations
considered were self-similar differing only in coupling strengths, a fact
that can be exploited computationally. Obviously
the situation requires a deeper understanding. In particular
it would be desirable to see explicit field transformations
relating the different reformulations. These could be intrinsically
interesting as they promise to generalise the vertex operator
constructions that have proved so important in string theory and the
representation theory of infinite dimensional algebras.

It is intriguing that the relevant class of quantum field theory
is so close to physically realistic unified gauge theories. Maybe
nature is that way. But then the question as to the whereabouts
of the magnetic monopole becomes acute. Perhaps they really are too
 heavy to pair  produce or maybe there is yet another theoretical twist.
One promising scenario is that they simply vanish from sight by
condensing into the vacuum [15]. The beauty of this is that the resulting
condensate can have a very desirable consequence, 
namely the confinement of quarks, by means of a dual Meissner
effect [16].

Despite the above questions and uncertainties, 
the framework of ideas has found
fertile ground in superstring theory, mainly because of the dominant
role of supersymmetry there. The structure is even richer because
quantum consistency requires superstrings to live in space-times of
ten dimensions and the charge carriers have to be extended objects
of varying dimensions, called branes.

The final conclusion is happy. There is clearly much to learn
and it promises to involve new physics and new mathematics.

I would like to thank the organisers  for the opportunity to give this talk.
I have tried to emphasise the conceptual points and I realise this
has been at the expense of the technical details. More of these,
together with additional references to the original literature,
can be found in accounts I have given elsewhere [17]. I do apologise
to the experts for some oversimplifications, and in 
particular for a wayward factor of two.

\bigskip

\ni [1] P.A.M. Dirac: \lq\lq Quantum Theory of Emission and Absorption
of Radiation",  \prs {\bf A114} (1927) 243-265.

\noindent [2]  P.A.M. Dirac: \lq\lq Quantised singularities in the electromagnetic field",
\prs {\bf A133} (1931) 60-72.

\ni [3] C.N. Yang and R.L. Mills;\lq\lq Conservation of isotopic spin and isotopic gauge invariance",  {\it Phys Rev} {\bf 96} (1954) 191-195.

\ni R. Shaw, University of Cambridge thesis (1955),
{\it The problem of particle types and other 
contributions to the theory of elementary particles.}

\ni [4] P. Higgs: 
\lq\lq Spontaneous symmetry breakdown without massless bosons",
{\it Phys Rev} {\bf 145} (1966) 1156-1163.

\ni [5] H. Georgi and S.L. Glashow: \lq\lq Unified Weak and Electromagnetic
Interactions without Neutral Currents'',
 {\it  Phys. Rev. Lett. }{\bf 28} (1972) 1494-1497.

\ni [6] G. 't Hooft; \lq\lq Magnetic monopoles in unified gauge theories",
{\it Nucl Phys} {\bf B79} (1974) 276-284.

\ni A.M. Polyakov;   \lq\lq Particle spectrum in quantum field theory",
{\it JETP Lett} {\bf 20} (1974) 194-195.

\ni [7] E.P. Wigner; \lq\lq Unitary
 Representations of the Inhomogeneous Lorentz Group ", 
{\it Ann. Math.} {\bf 40} (1939) 149-204.

\ni [8] T.H.R. Skyrme;  \lq\lq
Particle states of a quantized meson field", 
 {\it Proc Roy Soc} {\bf A262} (1961) 237-245.

\ni [9] S. Mandelstam:   \lq\lq Soliton
operators for the quantized sine-Gordon equation",  {\it Phys. Rev.} {\bf D11} (1975) 3026-3030.

\noindent [10] C. Montonen and D.I. Olive: \lq\lq Magnetic monopoles as gauge
 particles?", \plb {\bf 72} (1977) 117-120.

\noindent [11] A. D'Adda, R. Horsley and P. Di Vecchia:
 \lq\lq Supersymmetric Monopoles and Dyons", 
 \plb  {\bf 76} (1978) 298-302,

\noindent E. Witten and D.I. Olive: \lq\lq Supersymmetry Algebras 
that Include Topological Charges", \plb {\bf 78} (1978) 97-101,

\noindent H. Osborn: \lq\lq Topological Charges for $N=4$
 Supersymmetric
Gauge Theories and \break
 Monopoles of Spin 1", \plb {\bf 83} (1979) 321-326.

\ni [12] A. Sen: 
\lq\lq Dyon-monopole bound states, self-dual harmonic
 forms on the multi-\break monopole moduli space, and $SL(2,Z\!\!\!Z)$ 
invariance in string theory", {\it Phys Lett} {\bf  329B} (1994) 217-221.

\noindent [13] E. Witten: 
 \lq\lq Dyons of charge $e\theta/2\pi$", \plb {\bf 86} (1979) 283-287.

\noindent [14] H.A. Kramers and G.H. Wannier: \lq\lq Statistics of the two-dimensional ferromagnet I"
{\it Phys. Rev} {\bf 60} (1941) 252-276.

\ni [15] N. Seiberg and E. Witten: 
\lq\lq Electromagnetic duality, monopole condensation
 and confinement in $N=2$ supersymmetric Yang-Mills theory",
{\it Nucl Phys} {\bf B426}
 (1994) 19-52, {\it Erratum} {\bf B430} (1994) 485-486.

\ni [16] G. 't Hooft:
 \lq\lq Topology of the gauge condition and new confinement
 phases in nonabelian gauge theories",
  {\it Nucl Phys} {\bf B190} (1981) 455-478.

\noindent [17] P. Goddard and D.I.Olive: \lq\lq Magnetic Monopoles in Gauge Field Theories",
{\it Rep. Prog. in Phys.} {\bf 41} (1978) 1357-1437,

\ni D.I. Olive: \lq\lq  Exact Electromagnetic Duality",
 \npb (Proc Suppl){\bf 45A} (1996) 88-102,
\npb (Proc Suppl){\bf 46} (1996) 1-15,
\lq\lq Introduction to Duality", in
 \lq\lq Duality and Supersymmetric Theories", edited by D.I. Olive and P. West,
 (Cambridge University Press 1999),
62-94.

\bye